\documentclass[11pt]{article}

\pdfoutput=1

\usepackage{color,graphics,graphicx,enumerate,wrapfig,rotating}
\usepackage{array,epsfig,amsmath,amssymb,setspace,fullpage}
\usepackage{authblk}
\usepackage{amsthm}
\usepackage{subfigure}
\usepackage{url}
\usepackage[numbers]{natbib}
\usepackage{verbatim}

\usepackage{amsmath}
\usepackage{array}
\usepackage{tabularx}
\usepackage{todonotes}
\usepackage{changes}

\usepackage[noend]{algpseudocode}
\algrenewcommand{\algorithmiccomment}[1]{\ \ {\color{black!50}\texttt{/*} #1 \texttt{*/}}}
\algnewcommand\algorithmicto{\textbf{to}}

\begin{document}
\sloppy

\interfootnotelinepenalty=10000



\title{Neural Reconstruction Integrity:  A metric for assessing the connectivity of reconstructed neural networks \thanks{This material is based upon work supported by the Office of the Director of National Intelligence (ODNI), Intelligence Advanced Research Projects Activity (IARPA), via IARPA Contract No. 2012-12050800010 under the MICrONS program. The views and conclusions contained herein are those of the authors and should not be interpreted as necessarily representing the official policies or endorsements, either expressed or implied, of the ODNI, IARPA, or the U.S. Government. The U.S. Government is authorized to reproduce and distribute reprints for Governmental purposes notwithstanding any copyright annotation therein.}}

\author[1]{Elizabeth P. Reilly\thanks{Elizabeth.Reilly@jhuapl.edu}}
\author[1]{Jeffrey S. Garretson}
\author[1]{William Gray Roncal}
\author[1]{Dean M. Kleissas}
\author[1]{Brock A. Wester}
\author[2]{Mark A. Chevillet}
\author[1]{Matthew J. Roos\thanks{Matthew.Roos@jhuapl.edu}}
\affil[1]{Johns Hopkins University Applied Physics Lab}
\affil[2]{Facebook (work done while at JHU/APL)}

\maketitle

\begin{abstract}
Neuroscientists are actively pursuing high-precision maps, or \textit{graphs} consisting of networks of neurons and connecting synapses in mammalian and non-mammalian brains.  Such graphs, when coupled with physiological and behavioral data, are likely to facilitate greater understanding of how circuits in these networks give rise to complex information processing capabilities.  Given that the automated or semi-automated methods required to achieve the acquisition of these graphs are still evolving, we develop a metric for measuring the performance of such methods by comparing their output with those generated by human annotators (``ground truth'' data).
Whereas classic metrics for comparing annotated neural tissue reconstructions generally do so at the voxel level, the metric proposed here measures the ``integrity'' of neurons based on the degree to which a collection of synaptic terminals belonging to a single neuron of the reconstruction can be matched to those of a single neuron in the ground truth data.  The metric is largely insensitive to small errors in segmentation and more directly measures accuracy of the generated brain graph.  It is our hope that use of the metric will facilitate the broader community's efforts to improve upon existing methods for acquiring brain graphs.
Herein we describe the metric in detail, provide demonstrative examples of the intuitive scores it generates, and apply it to a synthesized neural network with simulated reconstruction errors.
\end{abstract}

\section{Introduction}

Traditionally, reconstructions of neural tissue at the voxel level are obtained by imaging tissue slices, mosaicing and aligning these 2D digital slices to form a 3D volume of voxels, and labeling voxels with unique neuron and synapse identifiers \cite{Saalfeld2012,Takemura2013,Lee2016}.  If neuron and synapse relationships are annotated as well (e.g., the post-synaptic portion of synapse $i$ is found on neuron $j$) then a brain graph reconstruction can be derived from the annotated tissue reconstruction.  Herein we use the term \textit{annotate} to encompass both labeling of voxels and annotating neuron-synapse relationships.

Although trained individuals can generate annotated reconstructions with high accuracy, the labor involved cannot feasibly scale to the larger tissue volumes needed to provide informative graphs.  Based on the labor estimate from a recent reconstruction effort \cite{kasthuri2015} it would take roughly 30,000 people-years to manually annotate a 1 mm\textsuperscript{3} volume.  To annotate tissue reconstruction at such scales, researchers are developing automated or semi-automated methods \cite{knowles2016rhoananet,Funke2012,Nunez-Iglesias2013,Helmstaedter2011}. These methods cannot yet achieve human-level annotation performance however, and a variety of metrics have been developed to measure the accuracy of semi-automated reconstructions as compared to ``ground truth''\footnote{Given that even expert human annotators do not always agree as to the proper labeling of a voxel or object, \textit{gold standard} may serve as better terminology than \textit{ground truth}. However, we use \textit{ground truth} since that is the term commonly used in machine learning literature.  Errors in manual annotations are commented upon further in the Discussion section.} reconstructions that are manually generated. Classic reconstruction metrics such as the Rand Index \cite{Rand1971objective} and variations thereof operate at the voxel level -- penalizing reconstructions for which all voxels of a given object do not have a corresponding object in the ground truth data with a one-to-one voxel match.

While neuronal morphology almost certainly plays a role in neural processing (e.g., dendritic integration and compartmental processing) it is likely that a graph representation composed solely of vertices (representing whole neurons or reconstructed portions) and directed edges (representing directed synapses) is nonetheless sufficient to allow for a substantial increase in our understanding of brain networks and the manner in which they process information.  As such, there are disadvantages to limiting oneself to voxel-level reconstruction metrics given that many voxel-level errors (e.g., minor neuron segmentation errors) do not result in erroneous brain graph connections.  Additionally, there are reconstruction techniques that do no operate on images \cite{marblestone2014rosetta} and thus cannot be fairly compared with image based techniques using voxel-level measures.  We present the Neural Reconstruction Integrity (NRI) metric, which is designed to be sensitive to aspects of a reconstruction that relate to the underlying brain graph, while being insensitive to those that do not.  This method allows for a direct assessment of graph connections, which may be performed even when annotations are not available or not created, as with emerging sequencing methods \cite{marblestone2014rosetta}.

\section{Evaluation criteria}

The primary function of the NRI metric is to evaluate the degree to which an annotated reconstruction contains a brain graph that is an accurate reflection of the true brain graph.  In large part this implies an insensitivity to neuron segmentation errors that do not impact the brain graph.  However, additional metric qualities are desirable.

\begin{itemize}

\item\textit{Can operate on relatively small volumes of ground truth data:} One of the largest challenges of evaluating the accuracy of a reconstruction is that little ground truth data is available due to the extensive manual labor needed to generate it.  Typical graph similarity metrics are removed from consideration since the volume of ground truth data will be much smaller than that generated by semi-automated methods.  As a result, the evaluation metric should not strictly be a graph connectivity metric, but rather a proxy metric that measures reconstruction aspects critical for representing an accurate graph.

\item\textit{Applicable at various levels of granularity:} The metric should be flexible enough to evaluate reconstructions at various levels of granularity including single neurons, a small number of neurons or neuron fragments, or large, densely-annotated volumes. This allows one to compute the metric on a variety of types of ground truth data (e.g., sparsely annotated or densely annotated). In addition it allows one to evaluate the fidelity of spatially restricted regions throughout a reconstruction volume as well as identify whether inaccuracies are uniformly scattered across the volume or if they are concentrated at a few poorly reconstructed neurons.  Global evaluation (a single metric score computed from the annotation intersection of the reconstruction volume and the ground truth volume) would allow one to measure overall improvement of a reconstruction method across reconstruction iterations or compare between reconstruction methodologies.

\item\textit{Provides locally independent scores:}  An intuitive requirement is that if an entire neuron is ``ground truthed'' (manually annotated) and scored by the metric, this score should not change if additional neurons are subsequently ground truthed and the metric is then reapplied to the original neuron.  Similarly, if the metric is applied to a geometrically local region, the score should not change if a \textit{spatially disjoint} region of the volume is subsequently ground truthed and the original region is re-scored.  We highlight this requirement because we found that alternative metrics based on information theory failed to fulfill this criterion.

\item\textit{Scales well to larger reconstruction and ground truth volumes:}  Computation of the metric should be feasible even as the size of reconstruction and ground truth volume grow over time.  Both are expected to grow substantially in coming years thanks to improvements in data acquisition technologies and targeted efforts such as the Intelligence Advanced Research Projects Activity (IARPA) MICrONS program {\cite{MICrONS}}. Based on expected output under that program, an evaluation metric should be capable of being computed on reconstruction volumes containing billions of synapses and hundreds of thousands of neurons, at a minimum.

\item\textit{Provide intuitive scores:} Ideally scores should fall in a limited range such as $[0,1]$ and be intuitively commensurate with reconstruction errors.

\end{itemize}

\section{Previous work}

As our goal here is to assess the accuracy of a reconstruction as it pertains to the brain graph, metrics that only assess neuron segmentation are not sufficiently informative.  For example, the error-free path length \cite{Helmstaedter2011} measures the frequency of errors made during manual skeleton tracing.  It is defined as the total length of neuron skeleton divided by the number of errors made during tracing. The connectivity of a neuron is not considered in this measure, simply how well the skeleton of a neuron is reconstructed. 

Several existing methods of evaluation assess the voxel-level similarity of a reconstruction volume and a ground truth volume.  For example, the Rand Index \cite{Rand1971objective}, Adjusted Rand Index \cite{Hubert1985}, and Warping Index \cite{Jain2010} are often utilized as image segmentation error measures.  The Rand Index applied to annotated images is defined as the proportion of pairs of voxels that are paired in the same segment in both ground truth and the reconstruction. If both neurons and synapses are annotated, the Rand Index can correlate with brain graph accuracy in some cases.  Frequently, however, this scoring method can give results that are poor characterizations of the accuracy of the reconstructed brain graph.  For example, large groups of voxels may be mislabeled yet connectivity is unaffected (e.g., mislabeling many voxels at the edge of a large diameter synapse-free process). Conversely, only small groups of voxels maybe mislabeled yet connectivity is substantially disrupted (e.g., voxels across dendritic spines are mislabeled, resulting in orphaned synapses on spine heads).

A more recently adopted voxel-level metric is the variation of information \cite{Nunez-Iglesias2013, Plaza2014}. Variation of information is an information theoretic measure defined as
$$
VI(S,G)=H(G|S) + H(S|G)
$$
where $S$ is a reconstruction, $G$ is ground truth, and $H$ is the entropy function.  It is possible to apply variation of information to abstracted neuron-synapse relationship information (the same information utilized by the NRI) rather than directly to voxel information. In that case, the variation of information when applied to a fully annotated (both reconstruction and ground truth) neural network has a number of desirable properties.  However, there is not a simple, well-behaved way to define $VI$ for a single neuron.  The key dilemma is that the $H(S|G)$ term cannot naturally be broken down into elements that are relevant to a single ground truth neuron.

Another approach that is similar in spirit to NRI is a line graph-based Graph $f_1$ score \cite{grayroncal2014}.  This metric also evaluates connectivity by focusing on true positive, false positive and false negative pathways connecting synapses.  However, this metric was applied only to dense full volumes and undirected graphs and performance on error sub-types was not systematically evaluated.

More recently, the tolerant edit distance (TED) was proposed as a segmentation evaluation metric aimed at assessing topological correctness \cite{ted2016}.  The TED was used in the 2016 Medical Image Computing and Computer Assisted Intervention (MICCAI) challenge on Circuit Reconstruction from Electron Microscopy (CREMI) \cite{CREMI}.  The TED is calculated at the image level, yet aims to capture topological errors, specifically splits and merges.  Calculation of the TED requires solving an integer linear program (ILP), which selects the relabeling of one segmentation to minimize the number of splits and merges with respect to another segmentation.  By selecting a reasonable tolerance threshold, the TED can ensure that `tolerable' errors, or those which don't affect the topology of the circuit, are ignored in the error calculation.  One potential issue with the TED is that the proposed ILP may not be computationally tractable, though this often is not the case in practice. And while the TED's tolerance of segmentation errors is a desirable quality with regard to a metric that characterizes brain graph accuracy, the TED metric does not measure connectivity and thus cannot serve in this capacity independent of additional metrics.

\section{Neural Reconstruction Integrity}

\subsection{Definition}
We propose a new reconstruction metric called the Neural Reconstruction Integrity (NRI) metric.  The NRI is a single neuron metric, which can be extended to a local network (a subset of neurons from the network, or a geometrically restricted region) or a global network metric.  For a given ground truth neuron, we consider all synaptic terminals associated with the neuron\footnote{Throughout this article we use the term \textit{neuron} generically, with recognition that elements in the ground truth are likely to be fragments of neurons rather than whole neurons, and elements in the reconstruction may be neuron fragments, merged neurons, merged neuron fragments, or even something non-neuronal altogether. Use of terms such as neuron fragment or neuron element are sometimes used to draw attention to this fact.}.
Presynaptic and postsynaptic terminals are treated independently--that is, only the presynaptic or postsynaptic ``half'' of a synapse is associated with a given neuron (except in the case of an autapse, in which case both halves of the synapse would be associated with the same neuron). The NRI description below assumes that terminals in the reconstruction volume and the ground truth have already been matched. A proposed method for performing this matching is discussed in a subsequent section.

\begin{figure}[ht]
\centering
\includegraphics[scale =.75] {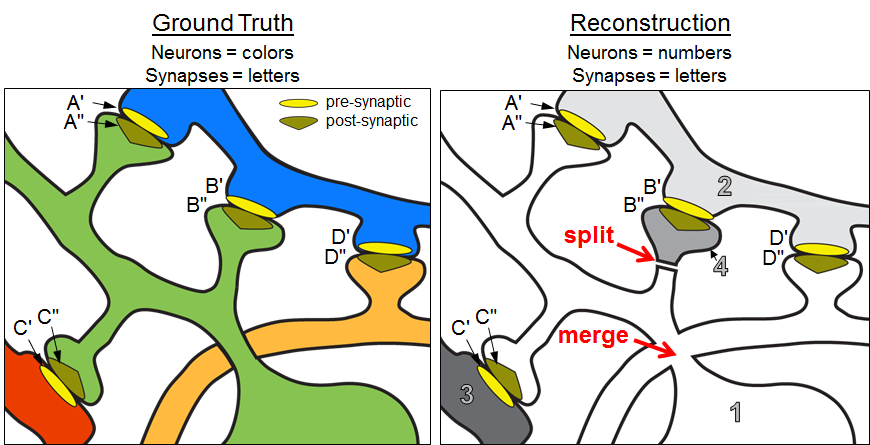}
\caption{\small{Ground truth neurons and a reconstruction containing split and merge errors. Focusing on the green neural fragment, A$''$-C$''$ is a true positive path, B$''$-A$''$ and B$''$-C$''$ are false negative paths, and D$''$-A$''$ and D$''$-C$''$ are false positive paths. The NRI score of the green ground truth neuron is 0.333 (based on the neural fragments and synaptic terminals shown in the panels). See text for additional details.}}
\label{fig:synPaths}
\end{figure}

The NRI measures the extent to which \textit{intracellular} paths between all possible pairings of ground truth synaptic terminals are preserved in the reconstruction.  For a pair of terminals on a ground truth neuron, a true positive indicates those two synaptic terminals are both associated with a single neuron in the reconstructed volume -- that is, an intracellular path is found between the terminals in the reconstruction.  For instance, in Figure \ref{fig:synPaths}, post-synaptic terminals A$''$ and C$''$ are correctly associated with the same neuron of the reconstruction, which yields a true positive.  However, B$''$ and C$''$ are not associated with the same neuron, yielding a false negative.

The NRI, then, is an $f_1$ score, which is the harmonic mean of precision and recall calculated on the true positive, false positive, and false negative paths as described above.  For a given ground truth neuron, $G_i$,
\begin{equation}\label{NRI_F1}
NRI(G_i) = 2\cdot\frac{precision \cdot recall}{precision + recall}
\end{equation}
where precision and recall have the usual definitions involving true positive (TP) counts, false positive (FP) counts, and false negative (FN) counts, $precision = \frac{TP}{TP+FP}$ and $recall=\frac{TP}{TP+FN}$.  Notice that, using the definitions of precision and recall, the NRI can be rewritten as
\begin{equation}\label{NRI_tpver}
NRI(G_i) = \frac{2\cdot TP}{2\cdot TP+FP+FN}
\end{equation}

To obtain a local network or global NRI value, one calculates the total number of TPs, FPs, and FNs over the set of ground truth neurons under consideration and uses these values to calculate the $f_1$ score as usual.  

Note that the global NRI value is strongly related to the line graph $f_1$ metric used in \cite{grayroncal2014}. In some sense, the NRI can be viewed as an extension of the line graph $f_1$, which also counts TPs, FPs, and FNs of intracellular paths in a reconstruction.  There are two key differences between the NRI and the line graph $f_1$ as defined and calculated in \cite{grayroncal2014}.
First, the NRI allows for evaluation at a variety of scales including single neurons, local networks, or global networks, allowing users to identify localized sources of error within the overall reconstruction in addition to achieving a snapshot performance of the entire network.  The second key difference is that the NRI operates on directed graphs, or a reconstruction where synapses have direction.  Accordingly, a neuron is penalized when one of its synapses is correctly identified in the reconstruction, but the direction is reversed -- a penalty that would not arise in the line graph $f_1$.  Despite these key differences we expect that, in many scenarios, the global NRI and the line graph $f_1$ would be highly correlated.

\subsection{Examples}
Consider Figure \ref{fig:synPaths} where a sample ground truth ``neuron'' (the green neuron) is reconstructed with a split error and a merge error.  In particular, a spine head (neuron 4 in the reconstruction) is split from the dendritic shaft of the neuron so the post-synaptic terminal B$''$ no longer has an intracellular path to A$''$ or C$''$.  This mistake yields two false negatives -- one for the lost A$''$ to B$''$ path and one for the lost C$''$ to B$''$ path.  Additionally, the orange neuron has been merged with the main body of the green neuron, resulting in new intracellular paths between D$''$ and the post-synaptic terminals A$''$ and C$''$.  The merged neuron element is labeled as 1 in the reconstruction. This merge yields two false positives -- one for the D$''$ to A$''$ path and one for the D$''$ to C$''$ path.  The intracellular path between A$''$ and C$''$ is retained, resulting in one true positive.  Using equation \ref{NRI_F1}, we obtain an NRI score of 0.333.

The NRI is degraded when neuron split, neuron merge, synapse insertion, and synapse deletion errors occur.  Synapse insertions increase the number of false positives while synapse deletions increase the number of false negatives.  Additionally, if the synapse direction is reversed, the NRI decreases due to additional false positives \textit{and} additional false negatives.  For example, in Figure~\ref{fig:synPaths}, if the presynaptic and postsynaptic terminals of synapse A were reversed so A$'$ was associated with neuron 1 and A$''$ was associated with neuron 2, then the NRI values of both the green and blue ground truth neurons decrease.  With respect to the green neuron, not only is the intracellular path between C$''$ and A$''$ absent (false negative), but a new path between C$''$ and A$'$ is introduced (false positive).

\subsection{Intuitive scores}

Here we highlight the intuitive relationship between reconstruction errors and the scores generated by the NRI metric. In each example scenario in Table~\ref{table:error_scenarios} it is assumed that all neurons have an equal number of synaptic terminals associated with them and that splits occur proportionately with regard to these terminals. We give global NRI scores (which are equal to single neuron scores in scenarios involving only one neuron) as well as precision (P) and recall (R).  Note that because NRI is a scalar metric its value does not indicate which types of reconstruction errors may have dominated in the event of a poor score.  However, low precision scores are solely due to neuron merges and synapse insertions, whereas low recall scores are solely due to neuron splits and synapse deletions.

\begin{table}
\begin{center}
\begin{tabular}{|  p{10cm}  ||  c  | c | c |}
\hline
\bf{Scenario} & \bf{P} & \bf{R} & \bf{Global NRI} \\
\hline
\hline
A neuron is split into two pieces with equal number of synapses & 1.00 & 0.50 & 0.67 \\
\hline
A neuron is split into three pieces with equal number of synapses & 1.00 & 0.33 & 0.50 \\
\hline
Two whole neurons are merged & 0.50 & 1.00 & 0.67 \\
\hline
Three whole neurons are merged & 0.33 & 1.00 & 0.50 \\
\hline
One neuron in a network of 10 neurons is split into 9 pieces and each piece is merged with one of the other 9 neurons & 0.82 & 0.91 & 0.86 \\
\hline
In a network of neurons, 20\% of synapses on each neuron are deleted & 1.00 & 0.64 & 0.78 \\
\hline
\end{tabular}
\caption{\small{NRI scores and the precision and recall components for various scenarios.  Scores are intuitively commensurate with the magnitudes and types of reconstruction errors.}}
\label{table:error_scenarios}
\end{center}
\end{table}

\section{Implementation of NRI}\label{section:implementation}

Computation of the NRI requires three steps: (1) pairing synapses in the ground truth with those in the reconstruction based on proximity, (2) summing the total number of matching synapses for every possible pair of ground truth neuron and reconstruction neuron, and assembling these sums into a \textit{count table}, and (3) using entries in the count table to determine the total number of true positive, false positive, and false negative pairs.

\subsection{Synapse alignment using centroids}

The first step is to determine which synapse(s) in the reconstruction correspond to synapses in the ground truth by synapse assignment, for which we propose using the Hungarian-Munkres algorithm \cite{kuhn1955hungarian} \cite{kuhn1956variants} \cite{munkres1957algorithms}. In general, assignment can be handled in a variety of ways depending on the format of existing data such as synapse centroids or labeled voxels.

In the following we assume that the information necessary for computing NRI has been extracted and stored in two data files -- one for the ground truth data and one for the reconstruction.  Each file contains a list of synapses with associated neurons and locations.  In other words, for a particular synapse the file contains an ID for the presynaptic neuron, an ID for the postsynaptic neuron, and an $(x,y,z)$ coordinate representing the centroid of the synapse.  There is no guarantee, and in fact it is unlikely, that the IDs or $(x,y,z)$ coordinates will correspond perfectly between the two lists due to reconstruction errors.  By applying the Hungarian-Munkres algorithm to synapse centroids, we reconcile the difference in synapse identifiers.  Note that it is not necessary to perform any neuron alignment, or any explicit pairing of ground truth neurons and reconstructed neurons.

Assigning synapses in the reconstruction to those in the ground truth can be nuanced, particularly if we consider volumetric synapse representations (labeled voxels).  For example, if the voxels of a reconstructed synapse overlap with half of those of a ground true synapse, and also overlap with an equal number of voxels outside of the ground truth synapse, it is somewhat subjective as to whether or not the reconstructed synapse should be assigned to the ground truth synapse.  However, the aim of the NRI metric is to measure characteristics important for representing brain graph connectivity rather than specific voxels or detailed synapse morphology. Thus, we propose the use of synapse centroids, which eliminates judgment calls based on the amount of voxel overlap.

Assigning synapses based on centroid locations still risks the introduction of assignment errors, however. Note that the centroid of a given ground truth synapse is unlikely to be perfectly matched with that of any from the reconstruction -- that is, centroid locations in the reconstruction can be viewed as being noisy estimates, and precise delineation of synapse boundaries is ambiguous. 

If ground truth synapses are dense in a particular region then assignment errors may occur when applying Hungarian-Munkres to the centroids. To ensure that the introduction of such errors has a negligible affect on the NRI score we simulated this assignment process by generating a synthetic distribution of synapses and adding location noise.  We modeled synaptic density as one synapse per cubic micrometer \cite{Schuz1989, anton2014three} and modeled centroid noise, synapse insertion rates, and synapse deletion rates based on data borrowed from Gray Roncal, et al. \cite{grayroncal2014}. Even at the highest ranges of noise, insertion rates, and deletion rates, the number of assignment errors was low -- approximately $5\%$ of the overall set of synapses.  To find an upper bound on the error of the precision measurement, we consider the worst case of overestimating false positives and false negatives by 5\% (calculation for recall is identical).  Denote false positives by FP and true positives by TP.  Assume $FP=r\cdot TP$ for $0\leq r\leq 1$ and we overestimate FP to be $1.05 \cdot FP$.  Then, using the definition of precision, our underestimate of the true precision can be rewritten as
\begin{equation}
Precision_{underest} = \frac{TP}{TP+1.05\cdot FP} = \frac{TP}{TP+1.05r\cdot TP} = \frac{1}{1+1.05r}.
\end{equation}
Similarly, the true value of precision is written as
\begin{equation}
Precision_{true} = \frac{1}{1+r}.
\end{equation}
Then, the overall error in our precision estimate is the difference of the two.
\begin{equation}
Precision_{error}=Precision_{true} - Precision_{underest} = \frac{1}{1+r} - \frac{1}{1+1.05r}.
\end{equation}
A plot of the error function for $0\leq r \leq 1$ shows this value is strictly less than 0.014.
Thus, an assignment error rate of 5\% will decrease the precision value for synapse detection by less than 0.014. Note that we also did not allow reconstruction synapses to be assigned to ground truth synapses with centroids further than $D$ away (e.g., $D=300$ nm), which is necessary to account for erroneous synapse deletions or insertions in the reconstruction.

\subsection{Count table calculation}

Once synapse assignment is complete, it is possible to generate the count table (a matrix).  In the count table, each row corresponds to a ground truth neuron and each column corresponds to a reconstructed neuron.  An entry in the table, $c_{ij}$, corresponds to the number of \textit{matched synaptic terminals} between ground truth neuron $G_i$ and reconstructed neuron segment $S_j$.
Matching synapse terminals are those for which both (1) the reconstruction synapse of neuron $S_j$ has been assigned to the ground truth synapse of neuron $G_i$, and (2) the polarity of the terminals are the same (presynaptic or postsynaptic).
Thus, if a terminal is presynaptic on $G_i$ in the ground truth and postsynaptic on $S_j$ in the reconstruction, then $G_i$ and $S_j$ do not share that terminal even though the synapses are assigned to each other.
Note that if $N$ reconstruction synapses are assigned to ground truth synapses, then there will be a total of $2N$ matching synaptic terminals in the count table (excluding those of the insertion row and deletion column -- see below). This applies for synaptic junctions with one pre-synaptic and one post-synaptic process, which is the case for the vast majority of known connections in mammalian cortex, but not for organisms such as \textit{drosophila}. Polysynaptic junctions will generate additional count table entries.



\begin{table}
\begin{center}
\begin{tabular}{| l | c | c | c | c | c |}
\hline
 & {\bf \textit{del}} & {\bf 1} & {\bf 2} & {\bf 3} & {\bf 4} \\ \hline
{\bf \textit{ins}} & 0 & 0 & 0 & 0 & 0 \\ \hline
{\bf green} & 0 & 2 & 0 & 0 & 1 \\ \hline
{\bf red} & 0 & 0 & 0 & 1 & 0  \\ \hline
{\bf blue} & 0 & 0 & 3 & 0 & 0  \\ \hline
{\bf orange} & 0 & 1 & 0 & 0 & 0  \\ \hline
\end{tabular}
\caption{\small{The count table for the ground truth and reconstruction depicted in Figure \ref{fig:synPaths}.}}
\label{table:counts}
\end{center}
\end{table}
The count table corresponding to Figure~\ref{fig:synPaths} is shown in Table~\ref{table:counts}. Examination of the count table immediately reveals useful information.  For instance, the ``green neuron'' was split into two elements in the reconstruction while ``neuron 1'' of the reconstruction is a merger of two ground truth neurons.

Additionally, the count table has a row corresponding to inserted synapses (\textit{ins}), or those found in the reconstruction and not the ground truth.  It also contains a column for deleted synapses (\textit{del}), or those found in the ground truth and not the reconstruction.

\subsection{Calculating NRI from the count table}

Once the count table is established, it is possible to calculate the NRI. For instance, notice that the number of true positives for the green ground truth neuron is $\binom{2}{2} + \binom{1}{2} = 1$, or the number of \textit{pairs} of green neuron terminals that are also found in the reconstruction\footnote{Where $\binom{n}{2}$ indicates $n$-choose-2, or the number of all possible pairs of elements from a set of $n$ elements.}.  The number of false negatives for the green neuron is $2\cdot 1=2$, or the number of pairs of terminals incorrectly split across neuron fragments in the reconstruction.  Finally, a false positive count may be obtained by looking at any given column.  For instance, the number of false positives associated with the green ground truth neuron is $(2\cdot 3)+(2\cdot 1) = 8$, which is then divided by two to prevent false positives from being double counted when they are summed over the entire network (see further explanation below).

Formally, let $C$ be the count table for a local network of the ground truth brain graph and the associated portions of the reconstruction.  The $0^{th}$ row refers to synapse/terminal insertions and $0^{th}$ column refers to synapse/terminal deletions while all other rows and columns indicate ground truth and reconstruction neurons, respectively.  There are $I$ total ground truth neurons and $J$ total corresponding reconstructed neurons (those that share at least one synapse with at least one ground truth neuron).  Neurons (or other objects such as glia) that share no synapse correspondences are ignored when computing NRI, as they do not impact our graph.  If $c_{ij}$ denotes the $i$,$j$-entry of the count table, then the total number of true positives, false negatives, and false positives across the volume can be computed using the equations below.

\bigskip
\noindent \textit{True positives:}

\begin{equation}\label{tps}
TP = \sum_{i=1}^I\sum_{j=1}^J \binom{c_{ij}}{2}
\end{equation}
Note that the outer summation is over the ground truth neuron index, $i$, thus the number of true positives for a single ground truth neuron is simply the inner summation over $j$ for a given $i$.

\bigskip
\noindent \textit{False negatives:}
\begin{equation}\label{fns}
FN = \sum_{i=1}^I  \Bigg[ \binom{c_{i0}}{2} + \sum_{j=0}^{J-1} \sum_{k=j+1}^J c_{ij}c_{ik} \Bigg]
\end{equation}
Notice that the false negative total includes contributions from the synapses in the deletions column (column 0) in two forms -- once with all synapses matched to those in the \textit{ground truth} neuron and again by pairing all possible combinations in the deleted column.  This ensures that the sum of the true positives and false negatives is equal to the total number of synapse pairs on the ground truth neuron. As for true positives, the number of false positives for a single ground truth neuron is simply the value of the term inside the inner summation, for a given neuron $i$.

\bigskip
\noindent \textit{False positives:}
\begin{equation}\label{fps}
FP=\sum_{j=1}^J  \Bigg[ \binom{c_{0j}}{2} + \sum_{i=0}^{I-1} \sum_{p=i+1}^I c_{ij}c_{pj} \Bigg]
\end{equation}
Computation of the total number of false positives is essentially identical to that for the false negative total, except computed in the other direction across the count table (effectively, computed on the transpose of the count table). Contributions from the insertions row (row 0) play a similar role to those from the deletions column under the false negatives computation -- being counted for incorrect pairing once with all synapses matches in the \textit{reconstructed} neuron and counted again for incorrect pairing in all possible combinations with each other.

Determining the number of false positives for a single ground truth neuron is open to interpretation, as there is ambiguity with regard to false positives that arise due to synapses being inserted on merged neurons. In addition, if two neurons are merged, the false positives created by the pairing of their synapses should be distributed between the neurons. In the latter case, we chose to attribute half the false positives to one neuron, and half to the other.  Regarding insertions, false positives due to pairs of inserted synapses are not attributed to a ground truth neuron (although false positives between an insertion and synapses found on a ground truth neuron \textit{are} attributed to that neuron) but they are added to the total count of network false positives.  Thus,

\begin{equation}\label{fps2}
FP = \sum_{i=0}^I FP(i)
\end{equation}
where $FP$ is the total count of network false positives, $FP(i)$ is the number of false positives attributed to individual ground truth neurons and (for $i=0$) those due to pairs of inserted synapses, and

\begin{equation}
  FP(i)=\begin{cases}
    \displaystyle \sum_{j=1}^J  \binom{c_{0j}}{2}, & \text{if $i=0$}\\[2.0em]
    \displaystyle \sum_{j=1}^J c_{ij}c_{0j} + \frac{1}{2}\sum_{j=1}^J \sum_{\substack{ p=1 \\ p\neq i}}^I c_{ij}c_{pj}    , & \text{otherwise}
  \end{cases}
\end{equation}


\bigskip
Once the total number of true positives, false positives, and false negatives have been tallied (for individual neurons or for the entire network), the final step is to use the calculated values in equation \ref{NRI_tpver} for a local network NRI value.

\section{Simulated data}\label{section:simdata}

To test the NRI metric behavior we would ideally apply it to a large 3D volume for which ground truth data existed, as well as semi-automated reconstructions generated over a range of methods and parameters. However, most currently available ground truth datasets tend to be small (hundreds of neurons) and sparse (few connections between neurons), and composed primarily of small fragments of neurons rather than large fragments or whole neurons, when compared to the volume of raw data currently being collected \cite{Takemura2013,Lee2016}.  We therefore chose to synthesize a neural network with modestly realistic anatomical properties, and introduce errors into the network (``perturb'' the network) to simulate reconstruction errors (resulting in imperfect reconstructions). This approach also allowed us to independently examine the effect of individual types of errors on the NRI scores, at graded perturbation levels.

To generate cortical networks with large numbers of neurons, we turned to NeuGen 2.0, a product developed at the University of Heidelberg, for generation of neurons and neural networks~\cite{eberhard2006neugen}.  NeuGen is an open source Java program that synthesizes neurons by using a probabilistic model of the growth of neuronal processes -- e.g., turning and branching. Processes are composed of numerous short, cylindrical segments. Synapse generation is based on Peter's Rule (distance between processes), modified to prevent synapse clustering (excessively dense synapse formation in localized process regions). Neurons were modeled after those in the rodent somatosensory barrel cortex as specified by the default NeuGen parameters. Our synthesized network consisted of 872 complete neurons (312 L2/3 pyramidal neurons, 62 L4 stellate neurons, 62 L4 star pyramidal neurons, 218 L5A pyramidal neurons, and 218 L5B pyramidal neurons) and over one million synapses -- approximately 2320 synaptic terminals per neuron, with somata confined in a volume of $x=y=79 \mu$m and $z=1300 \mu$m. Computational memory and processing limitations prevented us from generating a more dense network.  Although neuron density of the synthesized network is only about 1/10th that of real cortical tissue, we consider the network to be sufficiently large and complex to serve as a proxy for real data in testing of the NRI metric.

Current reconstruction methods generally introduce four types of reconstruction errors, with the error rates for each type often traded-off based on choice of algorithm parameters.  For example, synapse detection algorithms often have a tradeoff between synapse precision and recall, leading to added and/or deleted synapses in the final reconstruction.  Neuron segmentation algorithms may fail to differentiate membrane boundaries in poor quality images, resulting in merged neurons.  Yet if parameters are tuned to minimize false merges, the algorithm may identify nonexistent boundaries at thin portions of a neuron resulting in a neuron split (e.g., splitting of dendritic spines from the shaft).  To simulate the introduction of these errors into a reconstruction we built basic perturbation models for the generation of each type of error. Models are summarized in Table \ref{table:error_model}.

\begin{table}
\begin{center}
\begin{tabular}{| l || p{12cm} |} 
\hline
\bf{Error type} & \bf{Perturbation model description} \\
\hline
\hline
Synapse deletion & A specified percentage of synapses is randomly selected from the set of all existing synapses and deleted. \\
\hline
Synapse insertion & For each possible pair of cylindrical process segments (from different neurons), insert a synapse with probability $p$ where $p$ is $p_{max}$ for inter-process distance less than $d_1$, $p$ is 0 for distance greater than $d_2$, and $p$ follows a linear decreasing curve in $(d_1, d_2)$. \\
\hline
Neuron split & For each cylindrical process segment, split the neuron at the segment with probability $p$ where $p$ is $p_{max}$ for process diameter less than $d_1$, $p$ is 0 for diameter greater than $d_2$, and $p$ follows a linear decreasing curve in $(d_1, d_2)$. \\
\hline
Neuron merge & For each possible pair of cylindrical process segments (from different neurons), merge the neurons at the segments with probability $p$ where $p$ is $p_{max}$ for inter-process distance less than $d_1$, $p$ is 0 for distance greater than $d_2$, and $p$ follows a linear decreasing curve in $(d_1, d_2)$. \\
\hline
\end{tabular}
\caption{Descriptions of perturbation models used to produce imperfect graph reconstructions from a synthesized ground truth network.}
\label{table:error_model}
\end{center}
\end{table}

It is possible to run each perturbation model sequentially to generate all types of errors in a single reconstruction. However, in the following analysis, we generated reconstructions with only one type of error in each reconstruction, as this allowed direct observation of how the type of error affects neuron and network NRI scores.

\section{Applying the NRI to simulated data}

In this section, we empirically demonstrate relationships between error types and NRI values and give intuitive explanations of why these relationships exist.  The results in this section indicate that the NRI metric is well-behaved, scalable, and amenable to interpretation.  For each error type -- synapse deletion, synapse insertion, neuron split, and neuron merge -- the perturbation model is applied to the ground truth network described in Section \ref{section:simdata} with several different perturbation parameter sets, intended to create imperfect reconstructed graphs of decreasing accuracy (at the network level).  For example, in the case of synapse deletion, the percentage of synapses that are randomly deleted from the ground truth network is increased across individual simulations, resulting in reconstructed networks with different levels of synapse degradation.  Given the ground truth network and an imperfectly reconstructed network, the global NRI is calculated for the entire reconstructed network and the local NRI is calculated for each ground truth neuron.  Across the error types, we expect greater perturbation to lead to smaller NRI values.  This is the case for both local NRI (although scores vary from neuron to neuron) and global NRI.

\subsection{Synapse deletions and insertions}
\begin{figure}[!ht]
\centering
\includegraphics[scale =.35] {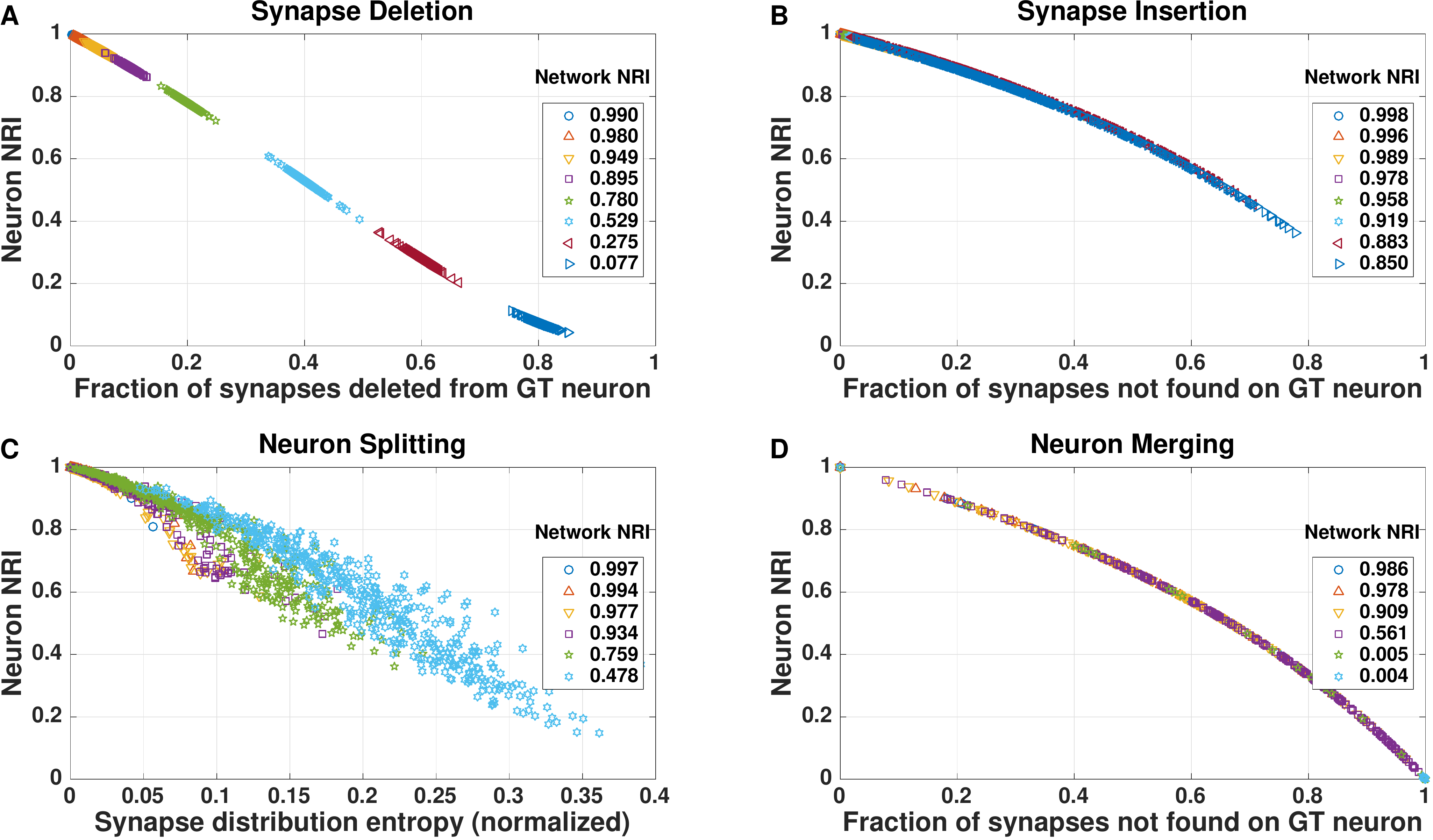}
\caption{\small{Reconstruction errors were simulated by applying one of four perturbation models to a synthetically generated ground truth (GT) network. Perturbation models (see Table~\ref{table:error_model}) introduced errors by (A) deleting a percentage of synapses from the overall network, (B) probabilistically inserting synapses where two neuron membranes are closely apposed, (C) probabilistically splitting neurons where process diameter is small, or (D) probabilistically merging neurons where two neuron membranes are closely apposed.  These plots show how local NRI scores of individual neurons vary as a result of the introduction of these errors.
Several perturbation metrics were used to compare perturbation magnitude to NRI scores. For synapse deletions, neuron NRI scores are compared to the fraction of synapses that were deleted from a given GT neuron. For synapse insertions and neuron merges, NRI is compared to the fraction of synapses \textit{not} found on the GT neuron. In the case of neuron merges, this means that if neuron $A$ is merged with neurons $B$ and $C$ in the reconstruction, then the perturbation score for neuron $A$ is $\frac{n_B+n_C}{n_A+n_B+n_C}$ where $n_A$, $n_B$, and $n_C$ are the number of synapses associated with neurons $A$, $B$, and $C$, respectively.
For neuron splits, neuron NRI is compared to the entropy of the synapse distribution across the split pieces of the GT neuron (normalized by the total number of synapses).  The color of each neuron's data point indicates the global network in which the neuron resided, and the NRI score for that global network is indicated in the plot's legend.  For example, for synapse deletions in plot A, the data points colored dark blue at the bottom right of the plot are neurons from a single perturbed network whose network NRI score is $0.077$.  Individual neuron NRI scores are close to the network NRI score in this particular case.
}
\label{fig:nri_plots}
}

\end{figure}

First, we consider synapse deletions.  As described in Table~\ref{table:error_model}, a fixed percentage of synapses are randomly chosen from across the entire volume and deleted.  Thus, most ground truth neurons will be impacted roughly to the same degree (with some variance about a mean).  When a single synapse is deleted, the number of true positives decreases and an equal number of false negatives is introduced.  The result is a lower recall score and a lower local NRI score.  The effect of decreased TPs and increased FNs is readily seen by studying equation~\ref{NRI_tpver}.  A synapse deletion only impacts the local NRI scores of the ground truth neurons with which the synapse is associated (presynaptic and postsynaptic).  The NRI decreases more for ground truth neurons that lose more synapses (as a fraction of total number of synapses associated with those neurons).  This is evident in Figure~\ref{fig:nri_plots} where the local NRI score is smaller for ground truth neurons that lose a greater fraction of their overall synapses.  Additionally, Figure~\ref{fig:nri_plots} shows that the network level or global NRI score also suffers when deletion rate is high.  For example, the dark blue markers in panel A represent individual neurons from a single reconstruction in which the deletion rate was high. Both the network and neuron NRI scores are low in this case.

Next, we consider synapse insertions.  Under the perturbation model, synapses are inserted probabilistically based on the distance between neuron membranes (more precisely, the distance between the cylindrical segments of which the neuronal processes are composed).  Naturally, some neurons will be significantly more impacted by this error model than others.  When a single synapse is inserted, several false positives are introduced where the number of false positives depends on how many synapses are associated with the original ground truth neuron.  False positives decrease the precision term and thus the total (local or global) NRI value.  Again, a synapse insertion effects the local NRI values of only the two neurons on which the synapse is incident (presynaptic and postsynaptic).
One measure of the extent to which a ground truth neuron has been impacted by insertions is the fraction of the reconstructed neuron's synapses that are not associated with those of the ground truth neuron.  This is the perturbation metric used in Figure~\ref{fig:nri_plots}B.
Neurons that experience a larger number of synapse insertions have lower NRI values, as seen in the figure.  Notice that, because this perturbation model will greatly impact a handful or neurons and leave others virtually untouched (due to the fact that the probability of insertion depends on the density of processes in the synthetic network, which is higher at the center of the volume and lower at the edges), Figure~\ref{fig:nri_plots}B does not show the same separation between reconstructed networks as Figure~\ref{fig:nri_plots}A does.  Global NRI values are not as heavily impacted and every reconstructed network has some neurons with low deletion and high NRI.

\subsection{Neuron splits and merges}

Segmentation errors made during reconstruction can result in neuron splits and neuron merges.  First, we consider neuron splits, which are made probabilistically based on process diameter (see Table~\ref{table:error_model}).  As with synapse insertions, the probabilistic model used will result in some neurons that are greatly affected by multiple splits and other neurons that are rarely or never split.  A single neuron split, say into pieces $A$ and $B$, will introduce several false negatives between all pairs of synapses where one synapse is associated with piece $A$ and the other synapse is associated with piece $B$.  Such an error only effects the NRI of the split neuron and the effect is immediately seen through inspection of equation~\ref{NRI_tpver}.  Figure~\ref{fig:nri_plots}C shows that greater splitting results in lower local NRI value.  Because neurons in a network are not uniformly impacted, there is no clear local NRI separation between neurons from low perturbation networks and those from high perturbation networks.


Finally, we consider neuron merges, which are made probabilistically when two neurons (processes) fall within a certain distance of each other.  Notice that, when this model is applied, whole neurons are merged together whenever a merge is indicated.  Thus, each ground truth neuron is a subset of a reconstructed neuron.
As for synapse insertions, we measure the extent to which a ground truth neuron has been impacted by merges as the fraction of the reconstructed neuron's synapses that are not associated with those of the ground truth neuron.  This is the perturbation metric used in Figure~\ref{fig:nri_plots}D.
Once again, the nature of the neuron merge model is that some neurons may be involved in several merges and others may be involved in a small number, possibly none.  Thus there is no clear separation in the NRI scores of high perturbation network neurons and low perturbation network neurons.  Merging two ground truth neurons, say $A$ and $B$, into one reconstructed neuron introduces a false positive for each synapse-synapse pair where one synapse is associated with neuron $A$ and the other is associated with neuron $B$ in the ground truth data.  The effect of additional false positives can readily be seen upon examination of equation~\ref{NRI_tpver}.  Figure~\ref{fig:nri_plots} verifies that ground truth neurons subject to a great deal of merging also tend to have small local NRI scores.

\section{Discussion}

\subsection{Simulation results}

Simulation results indicate that the NRI has several of the desired qualities of a metric for assessing reconstructions with regard to the brain graph accuracy.  For individual types of reconstruction errors, scores are intuitively commensurate with the magnitude of errors, with scores ranging from 0 to 1. Although not shown directly in the simulations (but see Table~\ref{table:error_scenarios}), when applied to reconstructions that contain multiple types of errors, observation of the precision and recall components of the NRI score lend additional insight into the types of errors contained in the reconstruction.  Finally, NRI computation was performed on a modern personal computer within run times on the order of seconds.  Although the simulated data sets were of modest size compared to that expected of real data sets in coming years, NRI computation on larger data sets will be feasible by utilizing the methods outlined in Section \ref{section:implementation} for synapse matching, and by leveraging more powerful computing hardware.

\subsection{Ground truth data}

We discuss here some aspects of real ground truth data that should be considered when applying the NRI metric.  Obtaining ground truth data through the manual sampling (annotating) of an image volume typically takes one of two forms -- densely annotating a geometrically confined region (e.g., a small cube within the larger volume) or sparsely annotating large portions of a few neurons and their processes, perhaps along with a subset of their synaptic partners.  In either case, we must remain aware that there is vastly more information in a large semi-automated reconstruction than in the ground truth data, and some aspects of the reconstruction may in fact be a more accurate depiction of the real brain graph than that depicted by the ground truth data.  

As a specific example, consider a branching process for which ground truth data exists for a pair of branches but not for the branching point (i.e., the branching point is outside of the manually annotated region). In this case, the ground truth data would label these processes as unique neuron fragments. However, if the larger reconstruction data captures the branching point, the two branches as well as the branching point would be correctly labeled as a unique neural fragment. If the NRI were computed on these data naively, the reconstruction would be unjustly penalized with many false positives since from the perspective of the ground truth data, the two branches were erroneously merged. Thus, a preprocessing step is needed in which the reconstruction is cropped to match the confined region of the ground truth data, and neuron fragments are relabeled based on connected components (i.e., generating two new identifiers for branches that do not have adjacent voxels in the cropped volume) such that cropped reconstruction labeling is equivalent to that which would have been obtained had the entire reconstruction been composed only of the confined ground truth region.

An additional problem arises when sparsely annotated ground truth data is used. In that case it is more likely that manual annotation errors will arise in the form of dendritic spine splits and associated orphaned synapses on spine heads, because all pixels are not assigned and so small details are more easily missed.
As mentioned in the introduction, ground truth should actually be treated as ``gold standard'' data, that, despite being used for assessing reconstruction quality, may itself have some errors.  One mitigating approach to the aforementioned problem is to revise the manner in which ground truth data is collected. For example, all synapses in the volume could first be annotated, and then traced back to a dendritic shaft, thereby reducing the likelihood of missing synapses.  Or as a compromise, the same approach could be taken but synapses would be annotated only within a fixed diameter range about a ground truth dendritic process, with the assumption that synapses outside this range could not belong to the dendrite. Finally, a modification to the NRI metric would make it insensitive to such errors, as described below.

\subsection{Future extensions}

In this manuscript, we defined an NRI operating point as the harmonic mean of precision and recall (e.g., $f_1$).  For graph inference tasks, it might be more favorable to choose a different $\beta$ value in $f_{beta}$, which has the effect of weighting the contribution of false positive and false negative paths asymmetrically.  Another extension would be to consider different methods of computing a global NRI score, such as weighting each neuron's contributions equally rather than weighted by the number of paths.  Many (brain)-graphs are produced without polarity information; NRI can be easily extended to undirected paths if desired.

\subsection{A modified, segmentation-only NRI}

Rigorous annotation methodologies are necessary to ensure that synapses are not missed when manually generating sparse ground truth annotations.  One approach to relaxing this requirement is to use a segmentation-only version of the NRI in conjunction with other metrics.  If the NRI is computed using only matched synapses (that is, unpaired synapses representing synapse deletions and synapse insertions are not included in the count table) then errors such as dendritic spine split errors in the ground truth data will not result in unjust penalization of the reconstructed neurons.

While this might appear to result in a metric that is insensitive to some errors in the reconstruction, this is only true if the associated spine synapses are deleted from the reconstruction as well.  In reality if the modified NRI is coupled with a synapse detection metric (as with the TED metric \cite{ted2016} in the 2016 MICCAI CREMI challenge \cite{CREMI}) and the score of the synapse detection metric is high, then spine segmentation quality will still be an important component of the NRI score.


\section{Conclusion}

We present an NRI metric for assessment of a reconstructed volume of neural tissue that emphasizes network connectivity. Our results indicate that the metric serves this purpose well based on several desirable qualities including applicability to both dense and sparsely annotated ground truth volumes, and applicability to single neurons, local regions, and global networks.  Additionally the metric produces an interpretable score that falls within $[0,1]$ and is computationally feasible even at scales much larger than that of currently available data sets.  We highlight NRI in the context of high-resolution brain graphs, but this metric applies broadly to graphs estimated using a variety of methods and at a variety of scales.  Indeed, it is potentially relevant for other problem domains where path finding is a critical objective (e.g., road detection, autonomy).

The metric has yet to be tested on a large volume of real ground truth data. In addition to confirming the utility of the metric, such an effort is likely to help refine strategies for manually annotating ground truth data and may ultimately facilitate researchers' efforts towards creating automated or semi-automated reconstruction methods leading to high quality, large scale brain graphs.



\nocite{*}
\bibliographystyle{IEEEtranN}
\bibliography{nri}

\end{document}